\documentclass[twocolumn,twoside,slac_two]{revtex4}
\usepackage{graphicx}
\usepackage{fancyhdr}
\begin{document}

\title{CHALLENGING THE HIGH ENERGY EMISSION ZONE IN FSRQs}


\author{A. Stamerra}
\affiliation{Dipartimento di Fisica, Universit\`a  di Siena, and INFN Pisa, I-53100 Siena, Italy}
\author{J. Becerra}
\affiliation{Inst. de Astrof\'{\i}sica de Canarias, E-38200 La Laguna, Tenerife, Spain}
\author{G. Bonnoli, L. Maraschi, F. Tavecchio}
\affiliation{INAF Osservatorio Astronomico di Brera, 23807 Merate, Italy}
\author{D. Mazin, K. Saito}
\affiliation{Max-Planck-Institut f\"ur Physik, D-80805 M\"unchen, Germany\\}

\author{on behalf of the MAGIC Collaboration}
\affiliation{\\}

\author{Y. Tanaka}
\affiliation{Institute of Space and Astronautical Science, JAXA, 3-1-1 Yoshinodai, Chuo-ku, Sagamihara, Kanagawa 252-5210, Japan}
\author{D. Wood}
\affiliation{Space Science Division, Naval Research Laboratory, Washington, DC 20375, USA\\}
\author{on behalf of the {\it Fermi}/LAT Collaboration}

\begin{abstract}
The blazar zone in quasars is commonly assumed to be located inside the broad-line region at some hundreds of Schwartzschild radii from the central black hole. Now, the simultaneous {\it Fermi}/LAT and MAGIC observations of a strong flare in the FSRQ PKS 1222+21 (4C 21.35, z=0.432) on 2010 June 17 challenge this picture. The spectrum can be described by a single power law with photon index 2.72$\pm$0.34 between 3\,GeV and 400\,GeV, and this is consistent with emission from a single component in the jet. The absence of a spectral cutoff constrains the gamma-ray emission region to lie outside of the broad-line region, which would otherwise absorb the VHE gamma-rays. On the other hand, the MAGIC measurement of a doubling time of about 10 minutes indicates an extremely compact emission region, in conflict with the "far dissipation" scenario. This could be a hint for the importance of jet sub--structures, such as filaments, reconnection zones or shear layers for the occurrence of blazar flares.

\end{abstract}

\maketitle

\thispagestyle{fancy}


\section{High energy emission from FSRQs}

The VHE (E$>100$ GeV) emission from the Flat Spectrum Radio Quasar (FSRQ) class of blazars has been assessed only recently with the present generation of Imaging Atmospheric Cherenkov Telescopes (IACTs), by the detection of 3C279 (z=0.536) \cite{MAGIC3c279} and subsequently  PKS1510-089 (z=0.36) \cite{1510HESS} and PKS1222+21 (a.k.a. 4C+21.35, z=0.432) \cite{MAGIC-ATel}. Among 50 extragalactic objects discovered in the VHE sky, the FSRQs are a minority, while the majority belongs to the BL Lac class of AGNs \footnote{For an updated list refer to http://tevcat.uchicago.edu/ or {http://www.mppmu.mpg.de/$\sim$rwagner/sources/} }.  On the contrary in the  high energy (HE) range, from 100 MeV to 100 GeV, FSRQs represent more than one third of the extragalactic objects. This has been confirmed with the {\it Fermi}/LAT observations as reported in the 2nd {\it Fermi}/LAT catalogue (see \cite{CavazzutiProc} these proceedings and \cite{LATAGN2ndCat}).

This gap  is motivated by the different convolution of the high--energy spectral emission of FSRQs, -- peaking in the hard-X/MeV bands --  with the different sensitivities of IACTs and orbiting $\gamma-$ray detectors. Moreover the absorption of VHE $\gamma$-rays by the extragalactic Background Light (EBL) affects the more distant FSRQs respect to the nearer BL Lacs. In any case the reduction of the energy threshold of IACTs, now as low as 50\,GeV for MAGIC, and the increase of their sensitivities should reduce the gap in the future.

FSRQs are the most interesting $\gamma-$ray objects for studying the multiple effect of jet acceleration and its interaction with the surrounding ambient medium. 
In fact, the common wisdom in the "blazarland" is based on the blazar sequence \cite{Fossati1998} with FSRQs leading the sequence with high bolometric luminosity, high accretion rates, radiatively efficient disk,  showing a luminous broad line region (BLR) $L_{BLR}\sim 10^{46} $ erg/s $\equiv L_{46}$ ($\sim 10$\% of the Eddington luminosity \cite{Dai2007}) and exhibiting a spectral energy distribution (SED) with a synchrotron peak in the IR/optical band. On the other end of the sequence we find the High Peaked BL Lacs (HBL) with low luminosity, radiatively inefficient disks, absence of detectable BLR, and  a SED dominated by the UV/X-ray synchrotron peaks.
This dichotomy has been confirmed in the {\it Fermi}/LAT AGN sample, showing the so-called "blazar divide" in the gamma-flux vs gamma-index plane, possibly driven by a single parameter, the jet power \cite{blazadivide}, ultimately connected to the accretion rate efficiency of the disk.

Within this frame the gamma-ray emission is commonly explained through inverse Compton scattering of accelerated electrons with a target radiation field: the same synchrotron emission (self synchrotron Compton, SSC) for  HBLs or the optical/UV photon field  of the BLR for FSRQs (external Compton emission, EC). Simple single emission zone models are able to explain the observed SEDs of HBLs and FSRQs with a reasonable set of parameters.  Some strain is envisaged in these models, such as the higher Doppler factors foreseen by the model respect to radio observations, the difficulties encountered to fit some extreme HBLs such as 1ES 0229+200\cite{0229Tavecchio}, S5~0716+714\cite{S50716MAGIC}) or the SEDs obtained from strictly simultaneous multi-wavelength (MWL) data as e.g. in~\cite{MAGICSwiftMrk421}.   To overcome these problems new set of models have been created, often invoking double emission zones (e.g. spine-layer \cite{spineLayer, Weidinger2010}), multiple emission zones \cite{Marscher2010}, also \cite{MarscherProc} these proceedings), thus relaxing the constraints through the increase of the number of parameters. The endorsement of these models needs new and more detailed observations to avoid the critics of being simple numerical artifacts.

\begin{figure*}[tbh]
\centering
\includegraphics[width=150mm]{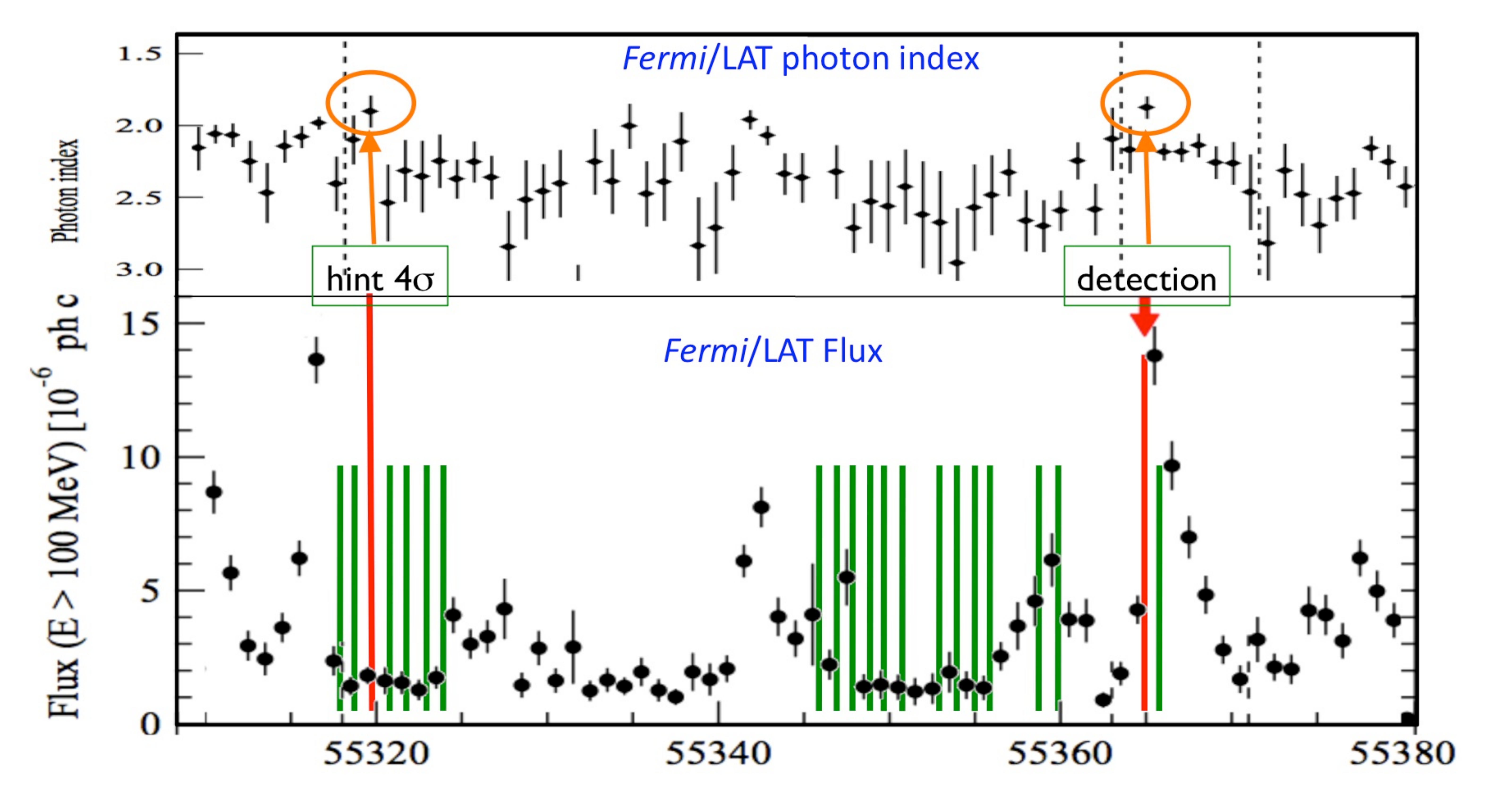}
\caption{ Lower panel: sequence of MAGIC observations (vertical bars) superimposed to the evolution of the $\gamma-$ray flux measured by {\it Fermi}/LAT during the observing campaign on PKS~1222+21 in flaring state. Upper panel: photon index in the HE range measured by {\it Fermi}/LAT; the lowest indexes, corresponding to MAGIC detections, are marked by two rings.} \label{fig:obs}
\end{figure*}

Although the scenario provided by the blazar sequence and the single zone emission models may not be so simple (\cite[e.g.][]{FossatiProc} these proceedings) it can justify the observational facts described -- double hump SED, gap in the ratio  of FSRQs and HBLs detected in the HE and VHE regimes -- and is a workbench where the jigsaw can lay to be unscrambled and deciphered.

In FSRQs the gamma-ray emission region, the "blazar zone" is  commonly located in the BLR, in the inner parsec region, according to the relation connecting the BLR size with its luminosity  $R_{BLR}\simeq 0.1\times L_{46}^{1/2}$~\cite{Kaspi2007}. The EC produced in the BLR can accommodate some observational facts such as the fast, daily, variability \cite{3C454.3} and the claimed GeV break in the HE spectra \cite{Poutanen2010}, but see also \cite{CostamanteProc} these proceedings.
Such "canonical scenario" for FSRQs also foresees a cutoff in the spectrum at few tenths of GeVs, due to $\gamma - \gamma$ absorption internal to the BLR~\cite{canonical}. This cutoff  can be shifted to higher energies $\sim~$TeV if the external radiation for the EC emission is provided by the IR torus at pc scales, or for lower compactness ratio of the BLR (see \cite{PoutanenProc}, these proceedings). 

The combined HE and VHE observations of PKS~1222+21 described in this work, will put severe constraints on the common scenario assumed for FSRQ unveiling more complex processes at work in the gamma-ray emission from these sources.

\subsection{PKS 1222+21}

PKS 1222+21 (a.k.a 4C +21.35, z=0.432) is a distorted quasar, showing a clear bended jet on VLBI images, at kpc distance and X-ray emission from the core and inner jet region \cite{JorstadMarscher2006}. Radio observations have measured superluminal motion (v$\sim 20c$ \cite{Homan2001}) in the inner jet region. Recent VLBA monitoring in the mm-wave band suggests a possible correlation between $\gamma-$ray flares and ejection of superluminal component from the mm-wave core (see \cite{JorstadBonn2010} and \cite{JorstadProc} these proceedings).
PKS\,1222+21 is a $\gamma$-ray blazar \citep{Abdo1LBAS} with a relatively hard spectrum in the GeV range. MAGIC detection \cite{MAGIC-ATel} and its inclusion in the list of $>$100~GeV emitters in the analysis of \citet{Neronov} makes it the third, and second most distant, FSRQ in the VHE band.

\section{Gamma-ray observations of PKS1222+21}

On 2010 April 24 the Astronomer Telegram 2584 \cite{ATel2584} reported an increase of the gamma-ray flux of 4C+21.35 measured by {\it Fermi}/LAT, exceeding of almost 10 times the alert threshold $10^{-6}$~ph/cm$^2$/s  at $E>100\,$MeV. The alert triggered many observatories in near IR~\cite{ATel2626}, optical polarimetry \cite{ATel2693}, X-ray \cite{ATel2698}, and were continued for almost two months following the high activity monitored by the LAT. An account of the MWL campaign was given during the symposium by Y. Tanaka \cite{TanakaProc}.  In figure \ref{fig:obs} the sequence of MAGIC observations superimposed to the LAT-monitored flux, is shown.

In the following the gamma-ray observations  performed by {\it Fermi}/LAT and MAGIC telescopes will be reported.
The complete analysis of the MWL campaign will be described in a forthcoming paper.

\begin{figure*}[hbt]
\centering
\includegraphics[width=110mm]{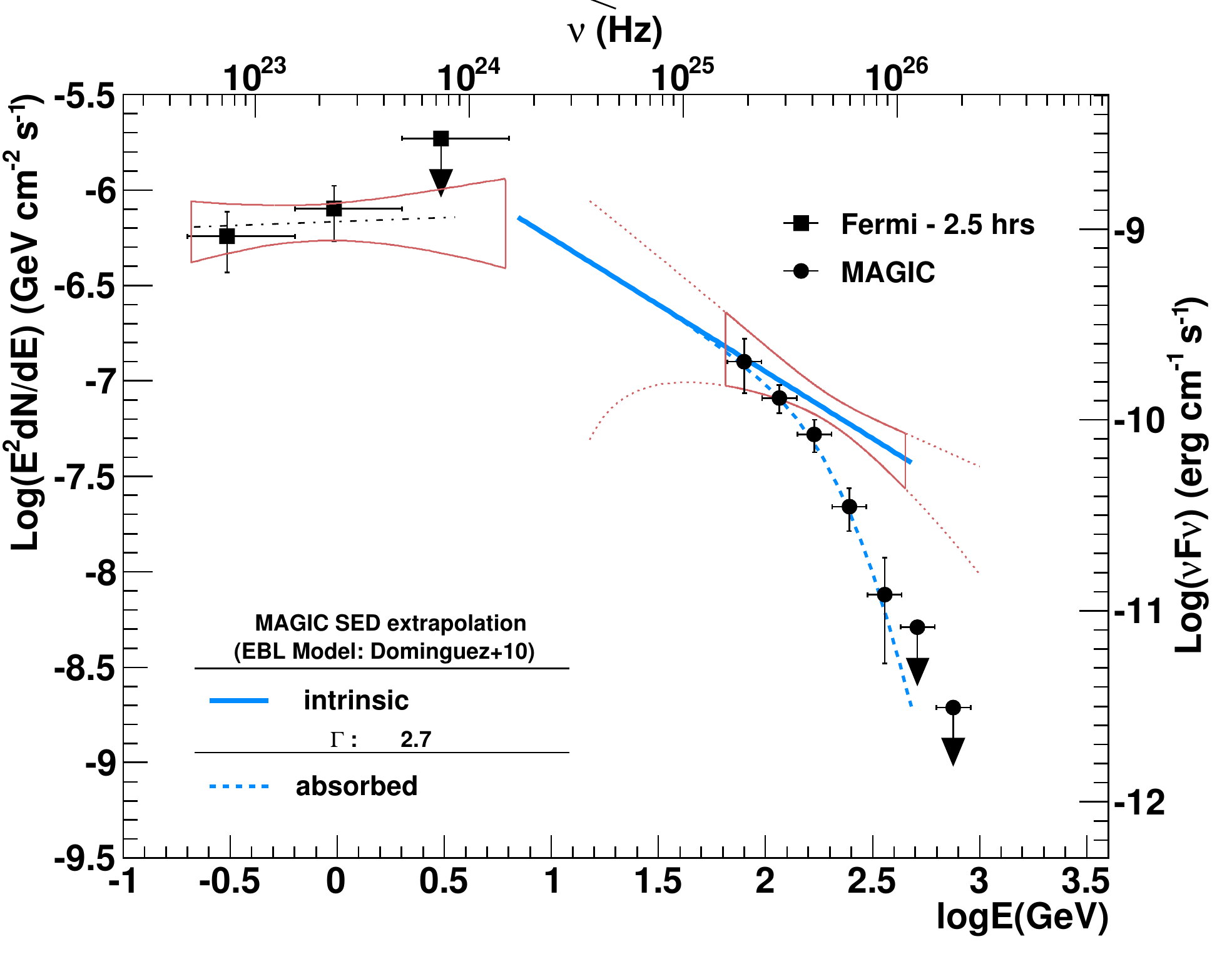}
\caption{High energy SED of PKS\,1222+21 during the flare of 2010 June 17 (MJD 55364), 
showing  {\it Fermi}/LAT (squares) and MAGIC (circles) differential fluxes. 
A red bow tie in the MeV/GeV range represents the uncertainty of the likelihood fit to the {\it Fermi}/LAT data.
The unfolded and deabsorbed spectral fit of the MAGIC data is also shown as a red bow tie, extrapolated to lower and higher energies (dotted lines). A thick solid line (photon index $\Gamma=2.7$) 
indicates a possible extrapolation of the MAGIC deabsorbed data to lower energies. 
The thick dashed line represents the EBL absorbed spectrum obtained from the extrapolated intrinsic spectrum using  the model  by \cite{Dominguez2010} }\label{SED}
\end{figure*}

\subsection{MAGIC observations}

The MAGIC telescopes system \footnote{http://wwwmagic.mppmu.mpg.de} observed PKS\,1222+21 between May $1^{st}$ and June $19^{th}$ 2010. A total amount of 16.5 hours of data was collected.  The details on the analysis can be found in the original discovery paper \cite{MAGICPKS1222}.

A hint of a signal was seen on May $3^{rd}$ at the level of  $4 \sigma$ significance. During 2.24 hours of observations an excess of $\approx$ 78 events was detected, amounting to 4.43 $\sigma$. MAGIC detected the VHE $\gamma$--ray emission from PKS\,1222+21 on June $17^{th}$ with statistical significance of 10.2 $\sigma$. The discovery was reported in \cite{MAGICPKS1222}. None of the other nights showed a significant excess of signal over background. 

% Interestingly the MAGIC detection and hint coincides with the hardest spectrum measured by LAT (see fig.\ref{fig:obs}), not corresponding to the flux peak during the outburst, as described by the clockwise evolution on the $\Gamma - (F>100\,$MeV) plane ({fig. 2(d) in}~ \cite{Tanaka2011}).

\subsection{{\it Fermi}/LAT observations}

{\it Fermi}/LAT detected the source in a active state in the period from April  to July 2010 with two prominent outbursts 
during the active state around MJD 55315 (2010 April 29) and 55365 (2010 June 18).
Details on the analysis and results for the active period can be found in \cite{Tanaka2011}.

We will concentrate on the outburst observed in coincidence with the MAGIC detection, when the source showed a significant flare lasting $\sim$3 days, with a flux peak on 2010 June 18. 
 A dedicated analysis found that the 1/2\ h  MAGIC observation fell
within a gap in the LAT exposure, thus we analyzed a period of  2.5\,h (MJD 55364.867 to 55364.973), encompassing  the MAGIC observation.
 The LAT analysis for this time bin was performed as in~\citet{Tanaka2011}, where
details can be found.

\section{The pieces of the jigsaw}

In figure \ref{fig:obs} the HE flux and photon index evolution is shown superimposed to the sequence of MAGIC observations. MAGIC observations did not result in a detection but two cases: a first hint on May 3rd and a firm detection on June 17th. We emphasize the correspondence of these two VHE signals with the lowest photon indexes measured by LAT, whereas the highest HE flux preceded or followed the highest flux  (see fig.\ref{fig:obs}).

Here the results on the MAGIC detection on 2010 June 17 and simultaneous LAT measurement will be outlined; details on the analysis can be found in the original paper \cite{MAGICPKS1222}.

\subsection{Spectrum and SED}

The energy spectrum measured by MAGIC spans from 70 GeV to 400 GeV and is well described by a simple power law form with a photon index $\Gamma=3.75 \pm 0.27_{stat} \pm 0.2_{syst}$ and a integral flux  $(4.6 \pm 0.5) \times 10^{-10} cm^{-2} s^{-1}$ at E$> 100\,$ GeV. The spectrum has been corrected for the Extragalactic Background Light (EBL) absorption, according to the model by \cite{Dominguez2010}, yielding an intrinsic spectrum with photon index $\Gamma_{intr} = 2.72 \pm 0.34$ between 70 and 400 GeV. The observed spectral points and the fit resulting from  the EBL correction are shown in the SED as a bow-tie in fig.\ref{SED}.
In the figure also the simultaneous spectral points measured by LAT and the resulting fit (bow-tie) are drawn. 
%They follow from a dedicated analysis in the time interval of 2.5 hours encompassing the 1/2 hour MAGIC observation. 
The simultaneous SED shows that the photon index steepens from 1.9 in the HE range to 2.7 in the VHE, and that there is a potentially smooth connection between {\it Fermi}/LAT and MAGIC extrapolated data in the 3 to 10 GeV region.
These results agree with the analysis of wider temporal intervals during this flare in which the source spectrum is well described by a broken power law with an energy break falling between 1 and 3 GeV \cite{Tanaka2011}.

\subsection{Lightcurve}

\begin{figure}[b]
\centering
\includegraphics[width=70mm]{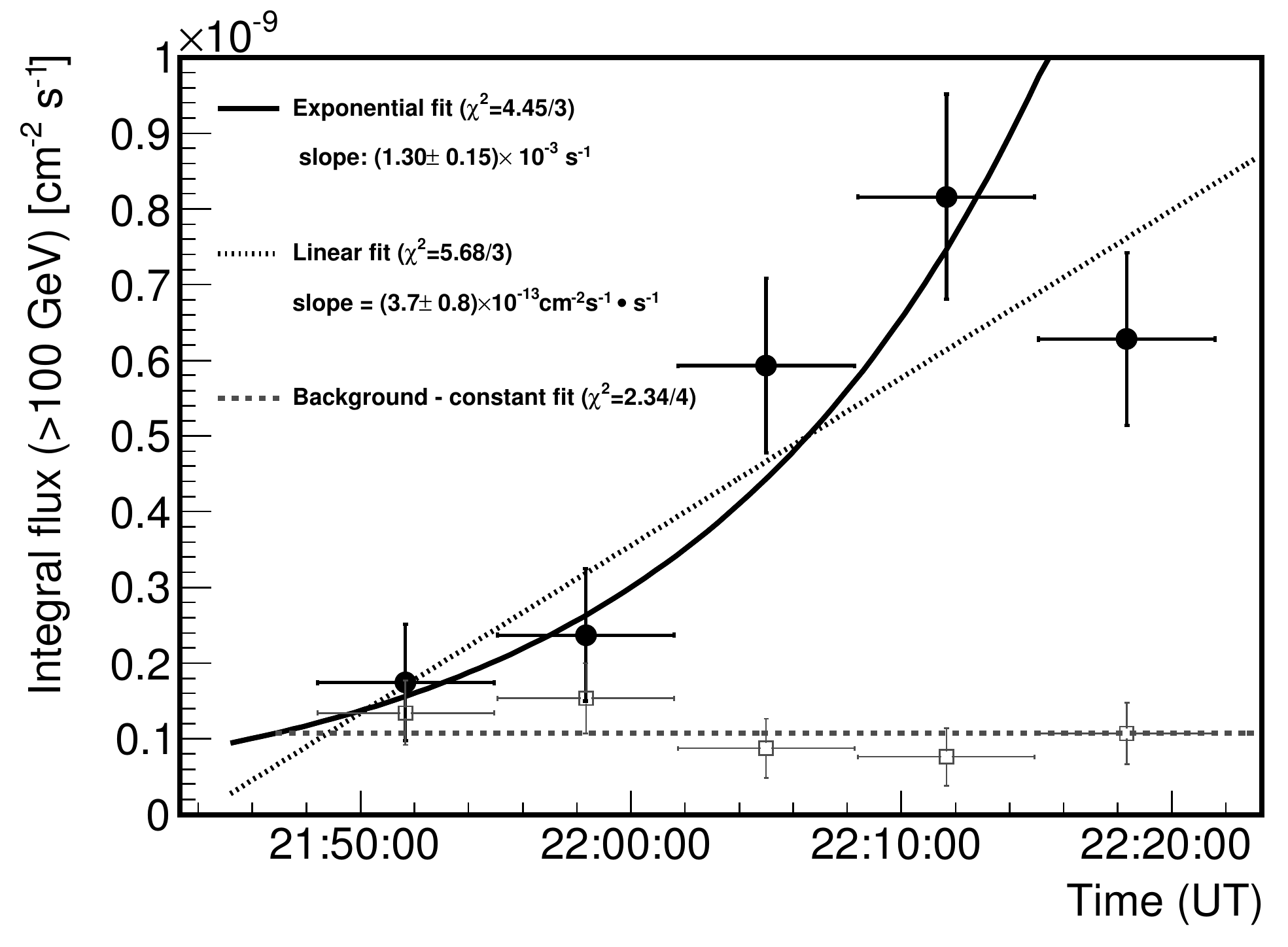}
\caption{PKS\,1222+21 light curve above 100\,GeV, in 6 minutes bins (black filled circles). The observation was carried out on MJD 55364.  The black solid line is a fit with an exponential function and the black dotted line a fit with a linear function. The grey open squares denote the fluxes from the background events and the grey dashed line is a fit with a constant function to these points.} \label{LC}
\end{figure}

The strength of the VHE emission allowed us to measure the flux in 6 minutes bins. The light curve in fig.\ref{LC} reveals clear flux variations, with a doubling time of $8.6^{+1.1}_{-0.9}$\, min, corresponding to the fastest time variation ever observed in a FSRQ in any energy range.

\section{The puzzle of the VHE $\gamma$-ray emission region}

The analysis of the combined LAT-MAGIC spectrum shows that data are fully compatible with a straight power law without cutoff in the observed VHE range; a deeper statistical investigation does provide a lower limit to the 
possible presence of cut-off in the VHE range to 130 GeV \cite{MAGICPKS1222}. Provided that further observations are needed to determine the expected steepening of the spectrum, this result challenges the standard scenario assuming the $\gamma-$ray emission region located within few Schwarzschild radii in the BLR. In fact,  a strong softening of the spectrum is expected above few tenths of GeV,  due to the opacity of the BLR to $\gamma-$rays and to the decreased efficiency of the IC scattering, see e.g.~\cite{canonical}. The importance of both effects  is reduced if the external photon field is associated with the IR torus, as envisioned by the "far dissipation" scenario \citep{Sikora2008}. In this case both effects start to be important above $\approx 1$\, TeV.

On the other hand the evidence of fast variability indicates an extremely compact  emission region with transverse dimensions, $R\sim 2.5\times 10^{14} (\delta/10) (t_{\rm var}/10 \,{\rm min})$ cm. This is difficult to be reconciled with the far dissipation scenarios if the emission takes place in the entire cross section of the jet, which is estimated to be $R\sim \theta _{\rm j} d \sim 3\times 10^{16} (\theta _{\rm j}/5 \, {\rm deg})$ cm ($d$: distance of the emitting region from the base of the jet $d>R_{\rm BLR}$;\,\,\,\,  $\theta_{\rm j}$: opening angle of the conical jet).

%\vspace{1mm}

\begin{figure*}[t]
\centering
\includegraphics[width=145mm]{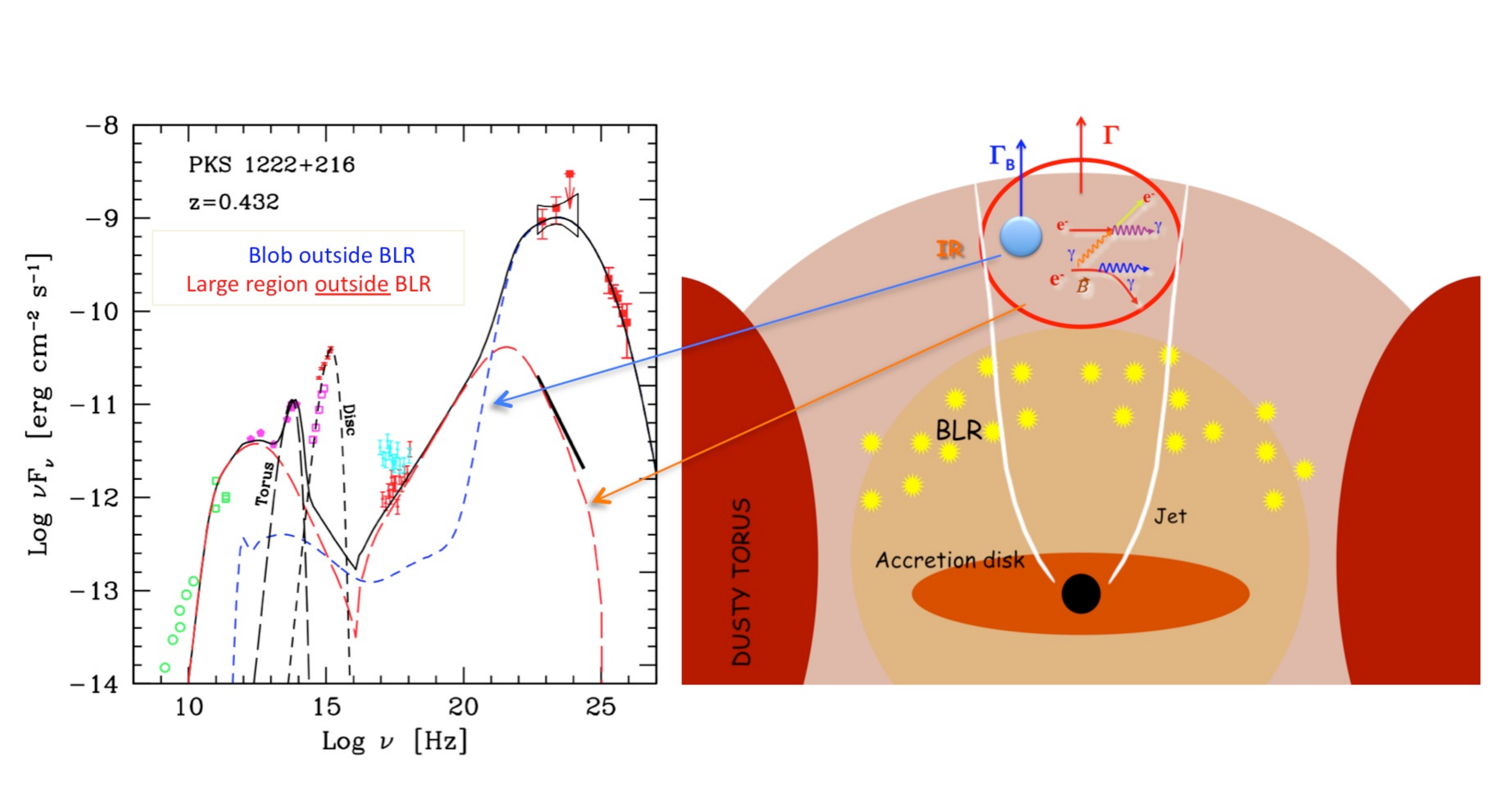}
\caption{Right: Sketch of the geometrical arrangement assumed in the model of double emission regions:  a small compact ÒblobÓ  with Lorentz factor $\Gamma_b$ and a underlying  emission  from a ÒstandardÓ spherical  region with radius R equal to the cross sectional size of a conical jet with semiÐ aperture angle, moving with bulk Lorentz factor $\Gamma$ located outside the BLR. See \cite{Becerra2011} for details. 
Left: Spectral energy distribution of PKS 1222+216 close to the epoch of the MAGIC detection (2010 June 17). Red points at optical--UV and X--ray frequencies. Fermi/LAT (red squares and Òbow tieÓ) and MAGIC data (corrected for absorption by the EBL) are the same shown in fig. \ref{SED}. The thick black solid line shows the LAT spectrum in quiescence (from \cite{Tanaka2011}). Magenta open squares are SDSS photometric points. Magenta filled pentagons are IR data. Green points report historical data.
The lines show the model corresponding to emission from the compact region and the standard jet. The blue short dashed line shows the emission from the compact region, while the red long dashed line reports the emission from the large region of the jet. The solid black line is the sum of the two. For details on the data and parameters of the models, refer to \cite{Becerra2011}.} \label{TwoZoneSED}
\end{figure*}

Some authors \cite{PoutanenProc} claim that the relation connecting luminosity and distance of the BLR region depends on too few measurements and may be more complex, thus reducing the compactness of the BLR and the corresponding $\gamma-$ray opacity. This would relax the constrain on the expected VHE cutoff.
Alternative scenarios pointing to emission beyond the BLR have been depicted, invoking the presence of very compact emission regions \citep{GhiselliniTavecchio2008, Giannios2009, Marscher2010} or through strong re-collimation mechanism, again forming small emitting nozzles e.g.~\cite{NalewajkoSikora2009, BrombergLevinson2009, Stawarz2006}.\\
As an example of such scenario in \cite{Becerra2011} the authors have developed a model where a compact emitting region, outside the BLR, embedded or not in the large-scale  jet, is responsible for the high energy emission and the larger region produces the lower energy emission, through synchrotron and IC or EC emission.
In figure \ref{TwoZoneSED} a sketch of the process and the resulting SED is shown. The resulting SED shows the feasibility of the observed flare in PKS1222+21 through a double zone scenario, though the physical mechanism producing such compact blob should be investigate further.

\section{Conclusions}

During this conference the discovery of the FSRQ PKS1222+21 by the MAGIC telescopes has been described. The simultaneous {\it Fermi}/LAT and MAGIC observations proved the $\gamma-$ray emission from 100 MeV to 400 GeV  posing tight constraints on the location of the emission region.
This observations prove that the present generation of IACTs has a high scientific potential when  promptly reacting to flaring states reported in the HE band ({\it Fermi}/LAT) and in the IR/optical band, associated to the BLR or to the IR-torus. Reduction of energy thresholds with future Cherenkov telescopes, such as CTA, will enlarge the gamma-ray horizon for such objects and together with a higher sensitivity should augment the number of new object of the FSRQ class in the VHE domain, thus placing new pieces for the completion of the still undiscovered puzzle of the location of the $\gamma-$ray emission region.

\bigskip % extra skip inserted
\begin{acknowledgments}

We would like to thank the Instituto de Astrof\'{\i}sica de
Canarias for the excellent working conditions at the
Observatorio del Roque de los Muchachos in La Palma.
The support of the German BMBF and MPG, the Italian INFN, 
the Swiss National Fund SNF, and the Spanish MICINN is 
gratefully acknowledged. This work was also supported by 
the Marie Curie program, by the CPAN CSD2007-00042 and MultiDark
CSD2009-00064 projects of the Spanish Consolider-Ingenio 2010
programme, by grant DO02-353 of the Bulgarian NSF, by grant 127740 of 
the Academy of Finland, by the YIP of the Helmholtz Gemeinschaft, 
by the DFG Cluster of Excellence ``Origin and Structure of the 
Universe", and by the Polish MNiSzW Grant N N203 390834.

The $Fermi$/LAT Collaboration acknowledges support from a number of
agencies and
institutes for both development and the operation of the LAT as well
as scientific
data analysis. These include NASA and DOE in the United States, CEA/Irfu and
IN2P3/CNRS in France, ASI and INFN in Italy, MEXT, KEK, and JAXA in
Japan, and the
K.~A.~Wallenberg Foundation, the Swedish Research Council and the
National Space
Board in Sweden. Additional support from INAF in Italy and CNES in France for
science analysis during the operations phase is also gratefully acknowledged.

This work was partially supported by DESY, a member of the Helmholtz Association (HGF).

\end{acknowledgments}

\bigskip % extra skip inserted

\begin{thebibliography}{99}   % Use for  1-9  references
%\begin{thebibliography}{99} % Use for 10-99 references

%\bibitem{accelconf-ref}
%http://www.cern.ch/accelconf

\bibitem{MAGIC3c279} Albert, J., et al. (MAGIC Collaboration) 2008a, Science, 320, 1752

\bibitem{1510HESS}Wagner, S., \& Behera, B.\ 2010, 10$^{\rm th}$
HEAD Meeting, Hawaii (BAAS, 42, 2, 07.05)

\bibitem{MAGIC-ATel}  Mariotti,~M., et al. (the MAGIC Collaboration),  Astronomer's Telegram \#2684 (2010)

\bibitem{CavazzutiProc} Cavazzuti E. ({\it Fermi} Collaboration), 2011, "Fermi and Blazars", these proceedings 

\bibitem{LATAGN2ndCat} Ackermann M. et al. \ (\textit{Fermi}/LAT Collaboration) 2011, Submitted ApJ, arXiv:1108.1420

\bibitem{Fossati1998} Fossati G. et al., 1998, MNRAS, 299, 433

\bibitem{Dai2007} H. Dai et al., 2007,  ApJ, 133, 2187

\bibitem{blazadivide} Ghisellini G., Maraschi L., Tavecchio F., 2009, MNRAS, 396, L105

\bibitem{MAGICSwiftMrk421} Aleksic J., et al. (MAGIC Collaboration), 2011, submitted ApJ, arXiv:1106.1589

\bibitem{0229Tavecchio} Tavecchio F., Ghisellini G., Ghirlanda G.,  Costamante L.,  Franceschin A.,  2009, MNRAS, 399, L59

\bibitem{S50716MAGIC} Anderhub H., et al. (MAGIC collaboration), 2009, ApJ, 704, L129

\bibitem{spineLayer} Tavecchio  F. \& Ghisellini G., 2009, MNRAS, 394, 131

\bibitem{Weidinger2010} Weidinger M., Spanier F. 2010, A\&A, 515, 18

\bibitem{Marscher2010} Marscher~A.~P. \& Jorstad~S.~G.\ 2010,  Proc. ``xFermi meets Jansky", arXiv:1005.5551

\bibitem{MarscherProc} Marscher~A.~P., 2011, "Observations and Modeling of Multi-waveband Variations of Blazars during $\gamma$-ray Outbursts", these proceedings

\bibitem{FossatiProc} Fossati G., 2011, "From sequence to envelope", these proceedings

\bibitem{Kaspi2007} Shai Kaspi et al., 2007, ApJ, 659, 997

\bibitem{3C454.3} Foschini L., Ghisellini G., Tavecchio F., Bonnoli G.,  Stamerra A., 2011, A\&A,  530, A77

\bibitem{Poutanen2010} Poutanen, J.\ \& Stern, B., 2010, ApJ, 717, L118

\bibitem{CostamanteProc} Costamante L., 2011, "Challenges from $\gamma$-ray Spectra of Blazars at
the Two Ends of the Blazar Sequence", these proceedings

\bibitem{canonical} Ghisellini,~G., \& Tavecchio,~F. 2009, MNRAS, 397, 985

\bibitem{PoutanenProc} Poutanen J., 2011, "Fermi observations of blazars and implication for the origin of gamma-rays", these proceedings

\bibitem{TanakaProc} Tanaka Y., 2011, "Fermi Large Area Telescope Detection of Bright Gamma-ray Outbursts from a TeV FSRQ 4C +21.35", these proceedings

\bibitem{Tanaka2011} Tanaka,~Y. T., et al.\ 2011, ApJ,  733, 19

\bibitem{Homan2001} Homan,~D.~C., et al.  2001, ApJ, 549, 840

\bibitem{JorstadMarscher2006} Jorstad,~S.~G., Marscher,~A.~P., 2006, Astron. Nachr, 327, 227

\bibitem{Abdo1LBAS} Abdo, A.\ A.\ et al. \ (\textit{Fermi}/LAT Collaboration) 2010b, ApJS,  183, 46

\bibitem{JorstadBonn2010}   Jorstad S.,  Marscher A., Smith P.,  Larionov V.,  Agudo I., 2010, Proc. ``xFermi meets Jansky"

\bibitem{JorstadProc} Jorstad S., 2011, "Parsec-Scale Jet Behavior of Blazars during High Gamma-Ray States", these proceedings


\bibitem{Neronov}
Neronov, A., Semikoz, D.\ V.\ \& Vovk, Ie.\ 2010, arxiv:1004.3767

\bibitem{ATel2584} Donato,~D.\ (\textit{Fermi}/LAT Collaboration)  2010, The Astronomer's Telegram, \#2584

\bibitem{ATel2626} Carrasco,~L.\ et al.  2010, The Astronomer's Telegram, \#2626

\bibitem{ATel2693} Dominici,~T.\ et al.  2010, The Astronomer's Telegram, \#2693

\bibitem{ATel2698} Verrecchia,~F.\ et al.  2010, The Astronomer's Telegram, \#2698

\bibitem{MAGICPKS1222}
Aleksi\'c, J., et al. (the MAGIC Collaboration), 2011, ApJL, 730, L8

\bibitem[Sikora et al.(2008)]{Sikora2008}
Sikora, M., Moderski, R., \& Madejski, G.~M.\ 2008, ApJ, 675, 71

\bibitem{Dominguez2010}
Dominguez, A.\ et al.\ 2011, MNRAS, 410, 2556

\bibitem{GhiselliniTavecchio2008} Ghisellini,~G., \& Tavecchio,~F.  2008, MNRAS, 386, 28

\bibitem{Giannios2009} Giannios,~ D., Uzdensky,~D.~A. ,Begelman,~M.~C. 2009, MNRAS, 395, 29

\bibitem{NalewajkoSikora2009}   Nalewajko,~K., Sikora,~M.\ 2009, MNRAS, 392, 1205

\bibitem{BrombergLevinson2009} Bromberg,~O., Levinson,~A. 2009, ApJ, 699, 1274

\bibitem{Stawarz2006}Stawarz,~L., et al.\ 2006, MNRAS, 370, 981

\bibitem{Becerra2011} Tavecchio F., et al., 2011, A\&A, 534, A86


%\bibitem{templates-ref}
%http://www.cern.ch/accelconf/templates.html

\end{thebibliography}

\end{document}